\documentclass{ifacconf}
\usepackage{amsmath} 
\usepackage{amssymb}  
\usepackage{algorithm}
\usepackage{algpseudocode}
\usepackage{siunitx}
\usepackage{graphicx}      
\usepackage{natbib}        
\begin{document}
\begin{frontmatter}
\title{R²NMPC: A Real-Time Reduced Robustified Nonlinear Model Predictive Control 
with Ellipsoidal Uncertainty Sets
for Autonomous Vehicle Motion Control} 


\author{Baha Zarrouki$^{*/**}$, João Nunes$^{*}$ and Johannes Betz$^{**}$}
\address{$^{*}$ Chair of Automotive Technology, Technical University Munich}
\address{$^{**}$ Professorship of Autonomous Vehicle Systems, TUM School of Engineering and Design, Technical University Munich, Munich Institute of Robotics and Machine Intelligence (MIRMI), \{{baha.zarrouki}, {joao.nunes}, {johannes.betz}\}@tum.de
}

\begin{abstract}                
In this paper, we present a novel Reduced Robustified NMPC (R²NMPC) algorithm that has the same complexity as an equivalent nominal NMPC while enhancing it with robustified constraints based on the dynamics of ellipsoidal uncertainty sets. 
This promises both a closed-loop- and constraint satisfaction performance equivalent to common Robustified NMPC approaches, while drastically reducing the computational complexity. The main idea lies in approximating the ellipsoidal uncertainty sets propagation over the prediction horizon with the system dynamics' sensitivities inferred from the last optimal control problem (OCP) solution, and similarly for the gradients to robustify the constraints. Thus, we do not require the decision variables related to the uncertainty propagation within the OCP, rendering it computationally tractable. Next, we illustrate the real-time control capabilities of our algorithm in handling a complex, high-dimensional, and highly nonlinear system, namely the trajectory following of an autonomous passenger vehicle modeled with a dynamic nonlinear single-track model.
Our experimental findings, alongside a comparative assessment against other Robust NMPC approaches, affirm the robustness of our method in effectively tracking an optimal racetrack trajectory while satisfying the nonlinear constraints. This performance is achieved while fully utilizing the vehicle's interface limits, even at high speeds of up to $37.5\unit{\meter\per\second}$, and successfully managing state estimation disturbances. Remarkably, our approach maintains a mean solving frequency of $144\unit{\hertz}$.
\end{abstract}


\end{frontmatter}
\section{Introduction}
Robust Nonlinear Model Predictive Control (NMPC) deals with uncertainty and disturbances in systems with nonlinear dynamics. Tube-based robust NMPC approaches maintain the system's state trajectories within a predefined "tube" despite disturbances (\cite{rawlings2017model}). The tube represents a region in state space where the system is guaranteed to stay, providing a safety margin against uncertainties. Some approaches employ central-path and ancillary controllers (\cite{mayne2011tube, yu2013tube, rubagotti2009robust}). Others enhance NMPC with incremental stability for constraint tightening using growing tubes (\cite{kohler_tube}).
Robust Min-Max NMPC approaches optimize control inputs by considering the worst-case scenarios for disturbances or uncertainties. It seeks to minimize the maximum possible cost or error over a range of disturbance scenarios (\cite{raimondo2009min, lazar2008input, chen1997game, chen2007disturbance}). However, it can be computationally expensive and can lead to overly cautious control actions due to its worst-case focus.\\
Our work draws inspiration from \cite{zanelli_zoro}, where the objective is to reduce the computational complexity associated with a robustified NMPC problem described in \cite{robust-mpc-original-paper}. The original problem involves incorporating ellipsoidal uncertainty sets into its optimization and propagating them alongside the nominal dynamics over the prediction horizon.
\cite{zanelli_zoro} employ an approximation approach by setting the first-order derivatives of the uncertainty dynamics to zero. This process enforces a sparsity structure within an inexact sequential quadratic programming algorithm. The outcome of this approximation is a reduction in the dimension of the optimal control problem (OCP), as the ellipsoidal sets are no longer treated as decision variables.
Nonetheless, this approach introduces a complex implementation that separates the solving process into two phases: a preparation phase that focuses on uncertainty propagation and the computation of approximate robustified constraints, followed by a subsequent feedback phase. This segmentation can potentially lead to computational times that exceed those of the original problem. It's essential to note that in the work of \cite{zanelli_zoro}, the evaluation does not encompass the computational time required for conditioning the solver with the propagated uncertainties and approximated constraints.
In our work, we also reduce the computational complexity of the robustified NMPC, making it equivalent to a nominal NMPC. This is accomplished by eliminating the decision variables associated with uncertainty propagation from the optimal control problem (OCP). Instead, we employ the sensitivities of the nominal dynamics, which are derived from the last NMPC solution spanning the entire prediction horizon, to approximate the propagation of ellipsoidal uncertainty sets for each shooting node. Notably, in contrast to \cite{zanelli_zoro}, our approach streamlines the OCP solving process into a single phase without the need for approximation of robustified constraints. Instead, we make use of the first-order derivatives of the original constraints from the previous solution to achieve robustification. 
In summary, this work presents three main contributions:
\begin{enumerate}
    \item We formulate a general Reduced Robustified NMPC (R²NMPC) incorporating robustified nonlinear constraints and uncertainty propagation with ellipsoidal dynamics while having the same complexity as an equivalent nominal NMPC.
    \item We showcase our R²NMPC’s closed-loop performance by applying it to control a high-dimensional and highly nonlinear system affected by severe disturbances —a full-scale autonomous vehicle.
    \item We present a thorough comparative analysis of our approach in relation to a Robustified NMPC and two other robust NMPC approaches from the state-of-the-art applied to the motion control use-case and demonstrate our R²NMPC robustness and real-
    time capabilities.   
\end{enumerate}
\label{sec:related_work}
\section{Nominal NMPC}
 \label{sec:nominal_nmpc}
We employ the following notation: Given a variable $z \in \mathbb{R}$, we define $\boldsymbol{z} = [z_0, z_1, \ldots, z_n]^T \in \mathbb{R}^n$ as the vector composed of $z$ variables. Additionally, $\boldsymbol{Z} = [\boldsymbol{z}^{(0)}, \boldsymbol{z}^{(1)}, \ldots, \boldsymbol{z}^{(m)}] \in \mathbb{R}^{n \times m}$ represents the matrix consisting of concatenated vectors $\boldsymbol{z}$.

We consider the following continuous time dynamical system in equation \ref{eq:dyn_sys}.
\begin{equation}
    \label{eq:dyn_sys}
    \dot{\boldsymbol{x}}(\tau) = f(\boldsymbol{x}(\tau), \boldsymbol{u}(\tau))
\end{equation} 
where $\boldsymbol{x} \in \mathbb{R}^{n_x}$ denotes the state vector, $\boldsymbol{u} \in \mathbb{R}^{n_u}$ the control vector, and $f$ the system dynamics. The respective nominal NMPC we consider for this system is given in equation \ref{eq:nominal NMPC problem} \cite{rawlings2017model}, with the respective cost function, equality and inequality constraints, and state propagation, 

\begin{equation}
\begin{aligned}
& \textbf{Problem 1} && \textbf{Nominal NMPC}\\ 
&  \underset{\boldsymbol{x}(.), \boldsymbol{u}(.)}{\min} & & 
\begin{aligned}
    \int^{T_p}_{\tau=0} & l(\boldsymbol{x}(\tau),\boldsymbol{u}(\tau)) \space  d\tau \\ 
 & + m(\boldsymbol{x}(T_p))
\end{aligned}
  \\ 
& \text{subject to} & & \boldsymbol{x}_{0} \leq \boldsymbol{x}(0) \leq \boldsymbol{x}_{0} \text {, }  \\
& & & \dot{\boldsymbol{x}}(t) = f(\boldsymbol{x}(t),\boldsymbol{u}(t)) \text {, } & t \in[0, T_p), \\
& & &  \boldsymbol{h}(\boldsymbol{x}(t), \boldsymbol{u}(t)) \leq \bar{\boldsymbol{h}},& t \in[0, T_p), \\
& & &   \boldsymbol{h}^{\mathrm{e}}(\boldsymbol{x}(T_p)) \leq \bar{\boldsymbol{h}}^{\mathrm{e}}, \\
\end{aligned}
\label{eq:nominal NMPC problem}
\end{equation}
where $t$ the discrete time, $T_p$ the prediction horizon, $\boldsymbol{h}$ and $\boldsymbol{h}^e$ the path and terminal inequality constraints on states and control inputs, which can be linear or nonlinear, and $\boldsymbol{x}_0$ the initial state.
Also, $l: \mathbb{R}^{n_{\mathrm{x}}} \times \mathbb{R}^{n_{\mathrm{u}}}  \rightarrow \mathbb{R}$ denotes the stage cost and $m: \mathbb{R}^{n_{\mathrm{x}}}  \rightarrow \mathbb{R}$ the terminal cost.

\section{Robust NMPC}
 \label{sec:robust_mpc}
The main idea of the Robust NMPC (RNMPC) we consider in this work is assuming ellipsoidal uncertainties and incorporating the dynamics of their propagation in the optimal control problem (OCP) formulation over the prediction horizon.
To this end, we consider the following disturbed dynamical system:
\begin{equation}
    \label{eq:dist_dyn_sys}
    \dot{\boldsymbol{x}}(\tau) = f(\boldsymbol{x}(\tau),\boldsymbol{u}(\tau),\boldsymbol{w}(\tau))
\end{equation}
where $\boldsymbol{w} \in \mathbb{R}^{n_w}$ denotes the disturbance vector. We also define an ellipsoidal set $\mathcal{E}(\boldsymbol{Q},\boldsymbol{q})$ as in Eq.\ref{ellipsoidal_set}:
\begin{equation}
    \label{ellipsoidal_set}
    \mathcal{E}(\boldsymbol{Q},\boldsymbol{q}) := \{\boldsymbol{q}+\boldsymbol{Q}^{\frac{1}{2}}\boldsymbol{v} | \exists \boldsymbol{v} \in \mathbb{R}^n : \boldsymbol{v}^T\boldsymbol{v} \leq 1 \}
\end{equation}
Here, $\boldsymbol{Q} \in \mathbb{S}_{++}^{n_w}$ is the set of symmetric positive definite matrices of dimension $n$. We model the disturbance $\boldsymbol{w}$ as part of an ellipsoidal set $\mathcal{E}(\boldsymbol{W})$, defined as \cite{phdthesis}:
\begin{equation}
    \label{eq:ellipsoidal_set_e}
    \mathcal{E}(\boldsymbol{W}) = \{\boldsymbol{W}^{\frac{1}{2}}\boldsymbol{v} | \exists \boldsymbol{v} \in \mathbb{R}^n : \boldsymbol{v}^T\boldsymbol{v} \leq 1 \}
\end{equation}
Where $\boldsymbol{W} \in \mathbb{R}^{n_w \times n_w}$ is the ellipsoidal disturbance matrix. \\
We approximate the true reachable set of the system's states 
(\cite{zanelli_zoro,phdthesis}). 
We define the initial set in which the state vector $\boldsymbol{x}$ is, by $\mathcal{E}(\boldsymbol{\Sigma}_0, \boldsymbol{x}_0)$ and an ellipsoidal from where we assume the disturbances come from $\mathcal{E}(\boldsymbol{W})$ (\cite{han2012lyapunov}).\\
At any time step $t$, we can write the state vector $\boldsymbol{\hat{x}}$ as the sum of a nominal state vector $\boldsymbol{{x}}$ with a deviation from the disturbance $\boldsymbol{\Delta x}$: 
\begin{equation}
\label{eq:nom_dist}
    \boldsymbol{x} = \boldsymbol{\hat{x}} + \boldsymbol{\Delta x}
\end{equation}
As $\boldsymbol{x}$ is not deterministic but affected by uncertainty $\boldsymbol{w}$, it can be described by its variance-covariance matrix $\boldsymbol{\Sigma}$, which defines our ellipsoidal uncertainty set. In the following, we formulate the ellipsoidal uncertainty dynamics equation used to propagate $\boldsymbol{\Sigma}$.\\
For sufficient small uncertainties, we separate the state dynamics $f$ in two parts: the nominal states 
and the deviations from the nominal trajectory 
(\cite{han2012lyapunov}). We also linearize the second term with first order derivatives (\cite{han2012lyapunov}), with $\boldsymbol{A}$ and $\boldsymbol{B}$ the Jacobians of the system dynamics (\cite{phdthesis}): 
\begin{equation}
\label{eq:sep_dyn}
    \boldsymbol{\dot{\hat{x}}} (\tau) = f(\boldsymbol{\hat{x}}(\tau),\boldsymbol{u}(\tau),\boldsymbol{w}(\tau))
\end{equation}
\begin{equation}
            \boldsymbol{\Delta{\dot{x}}} (\tau)= \boldsymbol{A}(\tau) \boldsymbol{\Delta x}(\tau) + \boldsymbol{B}(\tau)\boldsymbol{\Delta w}(\tau)
            \label{eq:sep_dyn_2}
\end{equation}
\begin{equation}
    \label{jacobians}
    \begin{aligned}
        \boldsymbol{A}(\boldsymbol{x}(t), \boldsymbol{u}(t)) := &\frac{\partial f(\boldsymbol{x}(t),\boldsymbol{u}(t), \boldsymbol{O}_{n_x})}{\partial {\boldsymbol{x}}} \\
        \boldsymbol{B}(\boldsymbol{x}(t), \boldsymbol{u}(t)) := &\frac{\partial f(\boldsymbol{x}(t),\boldsymbol{u}(t), \boldsymbol{O}_{n_x})}{\partial {\boldsymbol{w}}}
    \end{aligned}
\end{equation}

Since $\boldsymbol{\Delta x}(\tau)$ is a stochastic variable under the noise from $\boldsymbol{\Delta w}(\tau)$, we can calculate its variance-covariance matrix $\boldsymbol{\Sigma}(\tau)$ with Eq.\ref{eq: lyapunov_cont}. 
\begin{equation}
\label{eq: lyapunov_cont}
\begin{aligned}
    \boldsymbol{\Sigma}(\tau) = \mathbb{E}[\boldsymbol{\Delta x}(\tau)\boldsymbol{\Delta x}^T(\tau)]
\end{aligned}    
\end{equation}
Therefore, taking the first derivative in time from both sides, we formulate the 
propagation equation for the variance-covariance matrix:
\begin{equation}
    \label{eq: lyapunov_cont2}
    \frac{d\boldsymbol{\Sigma}(\tau)}{d \tau} = \lim_{\Delta \tau \Rightarrow 0} \mathbb{E}[\boldsymbol{\Delta x}(\tau+\Delta \tau)\boldsymbol{\Delta x}^T(\tau+\Delta \tau)]
\end{equation}
After converting the continuous-time system from Eq.\ref{eq:sep_dyn_2} to a discrete-time system with Euler Maruyama method, described in Eq. \ref{euler_mar} (\cite{lyapunov_equation_2}),
we can calculate the limit of the right-hand side of Eq.\ref{eq: lyapunov_cont2} and simplify it to the continuous Lyapunov Differential Equation (LDE) (\cite{behr2019solution}), in Eq.\ref{eq:lde} (\cite{lyapunov_equation_2}): 
\begin{equation}
\label{euler_mar}
    \boldsymbol{\Delta x}(\tau+\Delta \tau) = \boldsymbol{\Delta x}(\tau) +\Delta \tau \boldsymbol{\Delta x}(\tau)+\sqrt{\Delta \tau} \boldsymbol{\Delta w}(\tau) 
\end{equation}
\begin{equation}
    \label{eq:lde}
    \boldsymbol{\dot{\Sigma}}(\tau) = \boldsymbol{A}(\tau)\boldsymbol{\Sigma}(\tau) + \boldsymbol{\Sigma}(\tau) \boldsymbol{A}(\tau)^T + \boldsymbol{B}(\tau)\boldsymbol{W}(\tau)\boldsymbol{B}(\tau)^T
\end{equation}
Analog to Eq.\ref{jacobians}, the constraints can be approximated to:
\begin{equation}
\label{eq:inequality_constraints_correction}
    \boldsymbol{{h}}(\tau) = \boldsymbol{\hat{h}}(\boldsymbol{\hat{x}}(\tau), \boldsymbol{{u}}(\tau)) + \boldsymbol{C}(\tau) \boldsymbol{\Delta x}(\tau) 
\end{equation}
Here, $\boldsymbol{\hat{h}}$ denotes the nominal constraint and $\boldsymbol{C}$ is defined as:
\begin{equation}
\label{eq:jac_c}
    \boldsymbol{C}(\boldsymbol{x}(\tau), \boldsymbol{u}(\tau)) := \frac{\partial \boldsymbol{h}(\boldsymbol{x}(\tau),\boldsymbol{u}(\tau), \boldsymbol{O}_{n_x})}{\partial {\boldsymbol{x}}}
\end{equation}
The constraints $\boldsymbol{h}$ are stochastic (as $\boldsymbol{\Delta x}$ is stochastic). Following the demonstrations in \cite{han2012lyapunov}, we use the state transition matrix to estimate the worst case scenario of $\boldsymbol{w}$. With help of the Cauchy-Schwartz inequation, we calculate the new robustified inequality constraints (\cite{kurzhanski1997ellipsoidal, villanueva2017robust}):
\begin{equation}
    \label{eq:worst_case_ineq}
        \boldsymbol{{h}} = \boldsymbol{\hat{h}}(\boldsymbol{\hat{x}}(\tau), \boldsymbol{{u}}(\tau)) + \sqrt{\boldsymbol{C}(\boldsymbol{x}(\tau), \boldsymbol{u}(\tau))\boldsymbol{\Sigma}(\tau)\boldsymbol{C}^T(\boldsymbol{x}(\tau), \boldsymbol{u}(\tau))} 
\end{equation}
We formulate the general Robustified NMPC problem: \\
\begin{equation*}
\begin{aligned}
&\textbf{Problem 2} && \textbf{Robustified NMPC (RNMPC)}\\ 
&&&\textbf{with Ellipsoidal Uncertainty Sets}\\ 
\end{aligned}
\end{equation*}
\begin{equation}
\begin{aligned}
& \underset{\substack{\boldsymbol{x}(.), \\ \boldsymbol{\Sigma}(.), \\ \boldsymbol{u}(.)}}{\min}& & 
\begin{aligned}
    \int^{T_p}_{\tau=0} & l(\boldsymbol{x}(\tau),\boldsymbol{u}(\tau)) \space  d\tau \\ 
 & + m(\boldsymbol{x}(T_p))
\end{aligned}
  \\ 
& \text{s. t.} & & \boldsymbol{x}_{0} \leq \boldsymbol{x}(0) \leq \boldsymbol{x}_{0} \text {, }  \\
& & & \boldsymbol{\Sigma}_{0} \leq \boldsymbol{\Sigma}(0) \leq \boldsymbol{\Sigma}_{0} \text {, }  \\
& & & \dot{\boldsymbol{x}}(t) = f(\boldsymbol{x}(t),\boldsymbol{u}(t), \boldsymbol{O}_{n_x}) \text{, } & t \in[0, T_p), \\
& & & \dot{\boldsymbol{\Sigma}}(t) = \Phi(\boldsymbol{\Sigma}(t), \boldsymbol{W}(t), \boldsymbol{x}(t), \boldsymbol{u}(t)) \text{, } & t \in[0, T_p), \\
& & & 
\begin{aligned}
    \boldsymbol{h}&(\boldsymbol{x}(t), \boldsymbol{u}(t))  \\ & + \boldsymbol{h_{\text{back}}}(\boldsymbol{x}(t), \boldsymbol{u}(t), \boldsymbol{\Sigma}(t)) \leq \bar{\boldsymbol{h}}, 
   \end{aligned}
& t \in[0, T_p), \\
& & & 
\begin{aligned}
    \boldsymbol{h}^{\mathrm{e}}&(\boldsymbol{x}(T_p)) \\
    & + \boldsymbol{h_{\text{back}}}^{\mathrm{e}}(\boldsymbol{x}(T_p), \boldsymbol{\Sigma}(T_p))
\leq \bar{\boldsymbol{h}}^{\mathrm{e}}
\end{aligned}\\
\end{aligned}
\label{eq:original RNMPC problem}
\end{equation}
Here, $\boldsymbol{h_{\text{back}}}$ denotes the back-off term (\cite{zanelli_zoro} \cite{robust-mpc-original-paper}), i.e. robustification term through constraint tightening due to the uncertainty in the system:
\begin{equation}
    \label{eq:backoff_term}
    \begin{aligned}
         \boldsymbol{h_{\text{back}}}&(\boldsymbol{x}(t), \boldsymbol{u}(t), \boldsymbol{\Sigma}(t)) \\&:= \sqrt{\nabla_{\boldsymbol{x}} \boldsymbol{h}(\boldsymbol{x}(t), \boldsymbol{u}(t))^T\boldsymbol{\Sigma}(t) \nabla_{\boldsymbol{x}} \boldsymbol{h}(\boldsymbol{x}(t), \boldsymbol{u}(t))}
    \end{aligned}
\end{equation}
And $\Phi$ denotes the ellipsoidal uncertainty dynamics: 
\begin{equation}
    \label{eq:ellipsoidal_dynamics}
    \begin{aligned}
        \Phi&(\boldsymbol{\Sigma}(t), \boldsymbol{W}(t), \boldsymbol{x}(t), \boldsymbol{u}(t)) \\&:= \boldsymbol{A}(\boldsymbol{x}(t), \boldsymbol{u}(t)) \boldsymbol{\Sigma}(t)+\boldsymbol{\Sigma}(t)\boldsymbol{A}(\boldsymbol{x}(t), \boldsymbol{u}(t))^T\\
        & + \boldsymbol{B}(\boldsymbol{x}(t), \boldsymbol{u}(t)) \boldsymbol{W}(t)\boldsymbol{B}(\boldsymbol{x}(t), \boldsymbol{u}(t))^T
    \end{aligned}
\end{equation}

The computational complexity of the OCP formulated in Problem 2 renders it incompatible for online NMPC solutions and especially for real-time control applications: 
\begin{enumerate}
    \item This Robustified NMPC demands an ellipsoidal uncertainty dynamics propagation using $\Phi$ alongside the nominal dynamics using $f$ during the optimization problem and thus, requires a larger number of optimization variables than the nominal NMPC. The robustified OCP has a dimension of $n_x + \frac{n_x(n_x + 1)}{2}$.

    \item 
    The computation of the Jacobians $\boldsymbol{A}$ and $\boldsymbol{B}$ in Eq.\ref{jacobians}, i.e.
    the sensitivities of the nominal dynamics model $f$ w.r.t. system states and disturbances, to 
    approximate the ellipsoidal uncertainty dynamics $\Phi$ at each optimization stage is computationally expensive.

    \item Computing the back-off term in Eq.\ref{eq:backoff_term} for the robustified constraints at each optimization stage through the prediction horizon requires calculating the gradients of the nominal constraints w.r.t. the system states.
\end{enumerate}
To address these computational challenges, we propose a Reduced Robustified NMPC (R²NMPC) formulation:
\begin{equation*}
\begin{aligned}
&\textbf{Problem 3} && \textbf{R²NMPC with}\\ 
&&&\textbf{Ellipsoidal Uncertainty Sets}\
\end{aligned}
\end{equation*}
\begin{equation}
\begin{aligned}
&  \underset{\boldsymbol{x}(.), \boldsymbol{u}(.)}{\min} & & 
\begin{aligned}
    \int^{T_p}_{\tau=0} & l(\boldsymbol{x}(\tau),\boldsymbol{u}(\tau)) \space  d\tau \\ 
 & + m(\boldsymbol{x}(T_p))
\end{aligned}
  \\ 
& \text{s. t.} & & \boldsymbol{x}_{0} \leq \boldsymbol{x}(0) \leq \boldsymbol{x}_{0} \text {, }  \\
& & & \dot{\boldsymbol{x}}(t) = f(\boldsymbol{x}(t),\boldsymbol{u}(t), \boldsymbol{O}_{n_x}) \text{, } & t \in[0, T_p), \\
& & & 
\begin{aligned}
    \boldsymbol{h}&(\boldsymbol{x}(t), \boldsymbol{u}(t))  \\ & + \boldsymbol{h_{\text{back}}}(\boldsymbol{x}(t), \boldsymbol{u}(t), \boldsymbol{\Sigma}(t)) \leq \bar{\boldsymbol{h}}, 
   \end{aligned}
& t \in[0, T_p), \\
& & & 
\begin{aligned}
    \boldsymbol{h}^{\mathrm{e}}&(\boldsymbol{x}(T_p)) \\
    & + \boldsymbol{h_{\text{back}}}^{\mathrm{e}}(\boldsymbol{x}(T_p), \boldsymbol{\Sigma}(T_p))
\leq \bar{\boldsymbol{h}}^{\mathrm{e}}
\end{aligned}\\
\end{aligned}
\label{eq:R2NMPC problem}
\end{equation}
The main idea of the R²NMPC can be summarized in the following 3 differences to the original Robustified NMPC:
\begin{enumerate}
    \item We carry out the ellipsoidal uncertainty propagation outside the OCP. This simplifies the problem by eliminating optimization variables related to the propagation of ellipsoidal uncertainty sets $\boldsymbol{\Sigma}$ with $\Phi$, resulting in a reduced formulation that matches the dimension of the nominal NMPC, containing $n_x$ variables.
    \item We approximate the ellipsoidal uncertainty sets propagation $\dot{\boldsymbol{\Sigma}}$ using the sensitivities $\boldsymbol{A}$ and $\boldsymbol{B}$ (Eq.\ref{jacobians}) of the system dynamics obtained from the last R²NMPC solution at each shooting node with the discrete-time Lyapunov Dynamics, as in Eq.\ref{eq:discrete_Sigma_dynamics}. 
    This solution takes advantage of the feedback-loop and assumes that there is no big deviations in the dynamics between two consecutive R²NMPC solutions. 
    \begin{equation}
         \label{eq:discrete_Sigma_dynamics}
         \boldsymbol{\Sigma}(t+1) = \boldsymbol{A}(t)\boldsymbol{\Sigma}(t)\boldsymbol{A}^T(t)+\boldsymbol{B}(t)\boldsymbol{W}(t)\boldsymbol{B}^T(t)
     \end{equation}

    \item We approximate the back-off terms $\boldsymbol{h_{\text{back}}}$ at each shooting node using the gradients $\nabla_{\boldsymbol{x}} \boldsymbol{h}(\boldsymbol{x}(t), \boldsymbol{u}(t))$ computed from the last R²NMPC solution.
\end{enumerate}

Finally, Algorithm \ref{alg:RNMPC} illustrates the online implementation of our R²NMPC approach in a receding horizon fashion.
\begin{algorithm}
\caption{Implementation of the Reduced Robustified NMPC (R²NMPC)}\label{alg:RNMPC}
\begin{algorithmic}[1]
\State Set NMPC parameters: $T_p$, $T_{s}$, $Q$, $Q_e$ and $R$
\State Set R²NMPC specific parameters: $\boldsymbol{W}$ and $\boldsymbol{\Sigma_0}$
\For{$N_\text{sim}$ online steps}
    \State Update initial state $\boldsymbol{x}_0$ with current measurements
    \State Update current R²NMPC reference
    \State Solve the R²NMPC problem
    \State Apply the first control input $\boldsymbol{u}^*_0$ on the real system
    \For{$j \leq N_p$}
        \State Get current dynamics matrix $\boldsymbol{A}$
        \State Propagate the uncertainty matrix $\boldsymbol{\Sigma}$ ( Eq.\ref{eq:discrete_Sigma_dynamics})
        \State Calculate the constraints back-off terms (Eq.\ref{eq:backoff_term}) 
        \State update the constraints for stage $j$ 
    \EndFor
\EndFor
\end{algorithmic}
\end{algorithm}


\section{Reduced Robustified Nonlinear Model Predictive Control for the Trajectory Following of Autonomous Vehicles}
\label{sec:traj_following}
We tackle the task of following the trajectory of a full-scale autonomous vehicle by controlling its combined longitudinal and lateral motion. Our experimental platform is EDGAR, the TUM research vehicle (\cite{karle2023edgar}), a customized Volkswagen T7 Multivan tailored for autonomous driving software development. Our system encounters several types of disturbances, e.g. state estimation errors $\boldsymbol{x}_{t+1} = f(\boldsymbol{x}_{t},\boldsymbol{u},\boldsymbol{w}_t)$, where $\boldsymbol{w}_t$ denotes the uncertainties vector.
\subsection{Cost Function}
We establish the vehicle state vector as follows:
\begin{equation}
\begin{aligned}
\boldsymbol{x} &= [x_{\text{pos}},\space y_{\text{pos}},\space \psi,\space v_{\text{lon}},\space v_{\text{lat}},\space \dot{\psi},\space \delta_f,\space a]^T
\end{aligned}
\end{equation}
This vector encompasses various state components: $x_{\text{pos}}$ and $y_{\text{pos}}$ denote the ego vehicle's x and y coordinates, $\psi$ signifies the yaw angle, $v_{\text{lon}}$ and $v_{\text{lat}}$ represent longitudinal and lateral velocities, $\dot{\psi}$ corresponds to the yaw rate, $\delta_f$ indicates the front wheel's steering angle, and $a$ characterizes acceleration.
The control vector, denoted as $\boldsymbol{u}$, is defined as $\boldsymbol{u} = [j, \space \omega_f ]^T$, with $j$ representing longitudinal jerk and $\omega_f$ representing the steering rate at the front wheel.
For the cost function, we employ a nonlinear least squares approach for the stage cost: $l(\boldsymbol{x}, \boldsymbol{u})=\frac{1}{2}\|\boldsymbol{y}(\boldsymbol{x},\boldsymbol{u})-\boldsymbol{y}_{\mathrm{ref}}\|_W^2$, and for the terminal cost: $m(\boldsymbol{x})=\frac{1}{2}\|\boldsymbol{y}^e(\boldsymbol{x})-\boldsymbol{y}^e_{\mathrm{ref}}\|_{W^e}^2$. Here, $W$ and $W_e$ represent weighting matrices for the stage and terminal costs, respectively. Specifically, $W$ is computed as $W = \text{diag}(Q,R)$, where $Q$ and $R$ are matrices used for weighting states and inputs. $W_e$ is simply defined as $W_e = Q_e$.
\begin{equation}
  \begin{aligned}
\boldsymbol{y}(\boldsymbol{x},\boldsymbol{u}) &= [x_{\text{pos}},\space y_{\text{pos}},\space \psi,\space v_{\text{lon}},\space  j, \space \omega_f] \\
\boldsymbol{y}_{\text{ref}} &= [x_{\text{pos,ref}},\space y_{\text{pos,ref}},\space \psi_{\text{ref}},\space v_{\text{ref}},\space  0,\space 0] \\
\boldsymbol{y}^e_{\text{ref}} &= [x_{\text{pos,ref}}^e,\space y_{\text{pos,ref}}^e,\space \psi^e_{\text{ref}},\space v^e_{\text{ref}}]\\
\end{aligned}  
\end{equation}
To select the suitable matrices $Q$, $Q_e$, and $R$, we use Multi-Objective Bayesian Optimization, resulting in the following choices:
$
    Q = Q_e =  \text{diag} ( 
        2.8, \space 
        2.8, \space 
        0.4, \space 
        0.2
    )
$
and 
$
    R= \text{diag} ( 
        38.1, \space 
        101.4
    )
$.
\subsection{Nonlinear Prediction Model}
We employ a dynamic nonlinear single-track model that incorporates the Pacejka Magic Formula (\cite{pacejka1997magic}) to accurately represent crucial dynamic effects.
The system dynamics can be succinctly defined as: $f(\boldsymbol{x},\boldsymbol{u}) =$
\begin{equation}
\begin{aligned}
\begin{bmatrix}
v_\text{lon} \cos(\psi) - v_\text{lat} \sin(\psi) \space \\
v_\text{lon} \sin(\psi) + v_\text{lat} \cos(\psi) \space \\
\dot{\psi} \\
\frac{1}{m}\left(F_{x_r} - F_{y_f} \sin(\delta_f) + F_{x_f}  \cos(\delta_f) + m  v_\text{lat} \dot{\psi}\right) \\
\frac{1}{m}\left(F_{y_r} + F_{y_f} \cos(\delta_f) + F_{x_f} \sin(\delta_f) - m v_\text{lon} \dot{\psi}\right) \\
\frac{1}{I_z}\left(l_f \left(F_{y_f} \cos(\delta_f) + F_{x_f} \sin(\delta_f)\right) - l_r  F_{y_r}\right) \\
\omega_f \\
j  
\end{bmatrix} 
\end{aligned}
\end{equation}
Regarding the lateral forces $Fy_{\{f,r\}}$, we incorporate the combined slip effects of both lateral and longitudinal dynamics, following the methodology outlined in \cite{raji2022motion}. Here, the notation ${{f,r}}$ denotes either the front or rear tires.
\begin{equation}
\begin{aligned}
Fy_{\{f,r\}} &= F_{{\{f,r\},\text{tire}}} \cos\left(\arcsin\left(F_{x_{\{f,r\}}}/F_{max_{\{f,r\}}}\right)\right)
\end{aligned}
\end{equation}
To prevent issues related to singularities, we clip \\$F_{x_{\{f,r\}}}/F_{max_{\{f,r\}}}$ at 0.98, as in \cite{raji2022motion}. The lateral forces acting on the front and rear tires are computed using the simplified Pacejka magic formula formula (\cite{pacejka1997magic}):
\begin{equation}
\begin{aligned}
F_{{\{f,r\},\text{tire}}} &= D_{\{f,r\}} \sin(C_{\{f,r\}} \arctan(B_{\{f,r\}} \alpha_{\{f,r\}} \\
&- E_{\{f,r\}} (B_{\{f,r\}} \alpha_{\{f,r\}}- \arctan(B_{\{f,r\}} \alpha_{\{f,r\}}))))
\end{aligned}
\end{equation}
We define the side-slip angles as: 
\begin{equation}
\begin{aligned}
\alpha_f &= 
\delta_f - \arctan\left((v_\text{lat}+l_f \cdot \dot{\psi})/v_\text{lon}\right)  \\
\alpha_r &= 
\arctan\left((l_r \cdot \dot{\psi} - v_\text{lat}) /v_\text{lon}\right)
\end{aligned}
\end{equation}
The formula for the tire sideslip angle encounters a singularity problem concerning longitudinal velocity, as described in \cite{smith1995effects}. In order to resolve this, we make the assumption that the tire sideslip angle becomes negligible at low velocities, thus eliminating the necessity for two separate models. The definition of the longitudinal forces is as follows:
\begin{equation}
\begin{aligned} 
Fx_f &= - Fr_f \\
Fx_r &= F_d  - Fr_r  - F_{aero}\\
\end{aligned}
\end{equation}
In this context, the driving force applied at the wheel is characterized as $ F_d = m \cdot a$. Additionally, the rolling resistance forces are represented by 
$Fr_{\{f,r\}} = fr \cdot F_{z,{\{f,r\}}}$, where the rolling constant $f_r$ is specified as $fr = fr_0 + fr_1 \cdot \frac{v}{100} + fr_4 \cdot \left(\frac{v}{100}\right)^4$. Here, $v$ denotes the absolute velocity measured in $\unit{\kilo\meter\per\hour}$.
The aerodynamic force is computed using the formula: $F_{aero} = 0.5\cdot \rho \cdot S \cdot Cd \cdot v_\text{lon}^2 $\. Additionally, $F_{z,\{f/r\}}$ represents the vertical static tire load applied to both the front and rear axles. This is mathematically expressed as $F_{z,{\{f,r\}}} = \frac{m \cdot g \cdot l_{\{r,f\}}}{l_f + l_r}$. 
To determine the parameters of our prediction model, we conduct steady-state circular driving tests in accordance with ISO 4138 standards. For specific parameter values of the nonlinear single-track model and Pacejka tire model utilized in this study, please refer to Section V-I in \cite{karle2023edgar}.
\subsection{Constraints}
We express the potential limits on combined longitudinal and lateral acceleration for both R²NMPC (Problem 3) and the nominal NMPC (Problem 1) as a nonlinear constraint, as outlined in Eq. \ref{eq: acc potential}:
\begin{equation}
    h(\boldsymbol{x}, \boldsymbol{u}) = (a_\text{lon}/a_{x_{max}})^2 + (a_\text{lat}/a_{y_{max}})^2
    \label{eq: acc potential}
\end{equation}
In this context, the longitudinal acceleration is denoted as $a_\text{lon}$, which equals $a$, and the lateral acceleration is represented as $a_\text{lat}$, calculated as $v_\text{lon} \dot{\psi}$. The permissible upper and lower bounds for the variable $h$ are set as $\bar{h} = 1$ and $\underline{h} = 0$. We adjust the maximum allowable values to align with the constraints defined by the vehicle's actuator interface software. Specifically, we establish $a_{y_{max}} = 5.866 \unit{\meter\per\second\squared}$, while $a_{x_{max}}$ varies in accordance with the current velocity. When decelerating, the value of $a_{x_{max}}$ is determined as follows:
\begin{equation}
\begin{aligned}
a_{x_{max}} = 
\begin{cases}
     |-4.5\unit{\meter\per\second\squared}|,  &\text{if } 0\leq v_\text{lon} \leq 11 \unit{\meter\per\second} \\
     |-3.5\unit{\meter\per\second\squared}|,  &\text{if } 11 \unit{\meter\per\second} < v_\text{lon} \leq 37.5 \unit{\meter\per\second}
\end{cases}
\end{aligned}
\label{eq:deceleration constraints}
\end{equation}
And when the vehicle is in an acceleration phase:
\begin{equation}
\begin{aligned}
a_{x_{max}} = 
\begin{cases}
    3 \unit{\meter\per\second\squared}, & \text{if } 0\leq v_\text{lon} \leq 11 \unit{\meter\per\second} \\
     2.5\unit{\meter\per\second\squared}, & \text{if } 11 \unit{\meter\per\second} < v_\text{lon} \leq 37.5 \unit{\meter\per\second}
\end{cases}
\end{aligned}
\label{eq:acceleration constraints}
\end{equation}
Furthermore, we enforce strict linear constraints on the steering angle and the steering rate at the front wheel:
\begin{equation}
\begin{aligned}
-0.61 \unit{\radian}\leq & \delta_f \leq 0.61\unit{\radian} \\
-0.322\unit{\radian\per\second}\leq &\omega_f \leq 0.322\unit{\radian\per\second}
\end{aligned}
\label{eq:hard constraints}
\end{equation}

\subsection{R²NMPC configuration}
\label{subsec:R2NMPC_configuration}
For both our nominal NMPC and R²NMPC, we've configured the following parameters: a sampling time of $T_s = 0.08 \unit{\second}$ and a prediction horizon of $T_p = 3.04 \unit{\second}$. Specifically for R²NMPC, we work under the assumption that the system states, namely $v_\text{lon}$, $v_\text{lat}$, and $\dot{\psi}$, are influenced by bounded uniform disturbances. Our experiments have demonstrated that this assumption results in the most significant enhancements in closed-loop performance and computational efficiency. In the case of R²NMPC, we establish the following bounds on uncertainty in the form of a matrix:
\begin{equation}
\begin{aligned}
    \boldsymbol{W}^{\text{R²NMPC}} =& \text{diag}(\sigma_{vlon},\sigma_{vlat},\sigma_{\dot{\psi}}) \\ =& \text{diag}(1.1\unit{\meter\per\second}, 0.2\unit{\meter\per\second},0.05\unit{\radian\per\second})
\end{aligned}
\label{eq:uncertanty_bounds_RNMPC}
\end{equation} 
The choice of the $\boldsymbol{W}^{\text{R²NMPC}}$ matrix was based on the following considerations:
$\boldsymbol{\Sigma}_0 = \boldsymbol{\Sigma}(t=0)$
\begin{equation}
\begin{aligned}
    \boldsymbol{\Sigma_0}=& \text{diag}(\sigma_x,\sigma_y,\sigma_{\psi},\sigma_{vlon},\sigma_{vlat},\sigma_{\dot{\psi}},\sigma_{\delta_f}) \\ =& \text{diag}(0 ,0, 0, 0.6\unit{\meter\per\second}, \\ & 0.02   \unit{\meter\per\second},0.00125
\unit{\radian\per\second},0)
\end{aligned}
\label{eq:Sigma_0_RNMPC}
\end{equation} 
Here, $\boldsymbol{\Sigma_0}$ is the initial value of the uncertainty matrix. 

\section{Simulation Results}
\label{sec:sim_results}
In this section, we perform a performance evaluation by comparing our R²NMPC approach with an equivalent nominal NMPC. Both approaches are subjected to substantial additive disturbances that impact the state estimates. Additionally, we benchmark our approach against other state-of-the-art Robust NMPC methods applied to the motion control use-case.
\subsection{Simulation Setup}
\label{subsec:simulation_setup}
\begin{figure}[h]
\includegraphics[width=1\columnwidth]{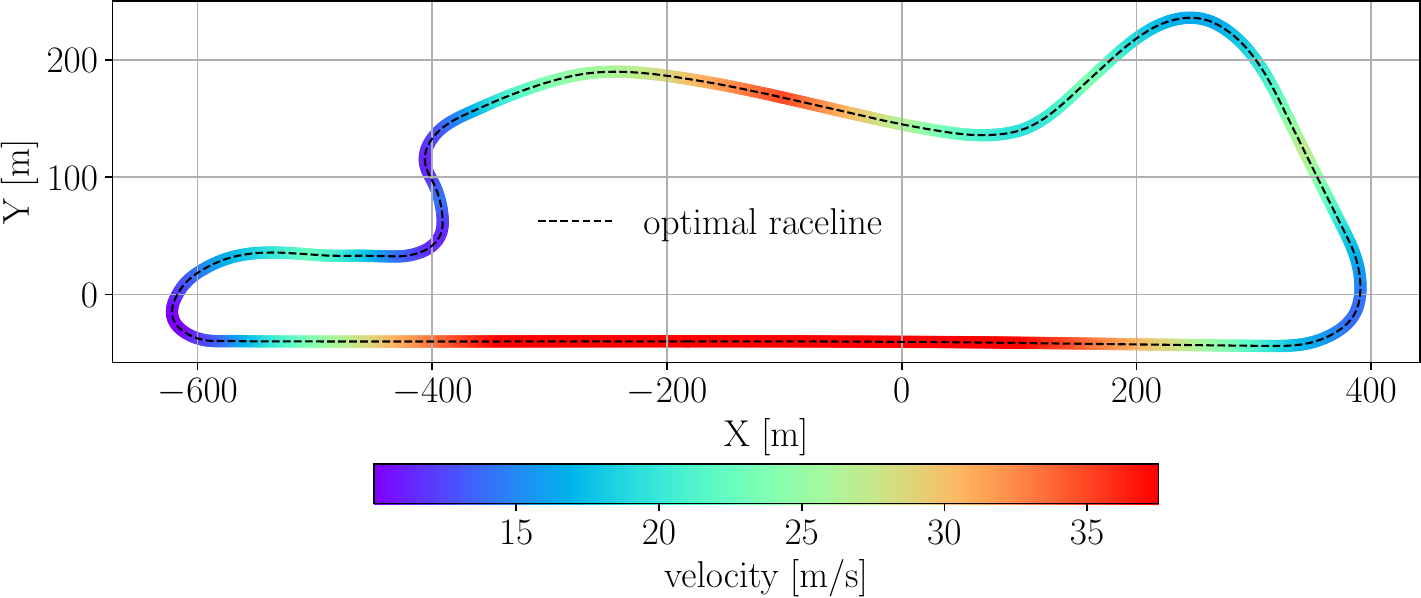}
\caption{Track layout and velocity profile at Monteblanco}
\label{fig:racetrack}
\end{figure}

Our experiments were carried out on a standard laptop equipped with an Intel i7-11850H 2.50GHz CPU and 16GB of RAM. The laptop was running Ubuntu 22.04 as the operating system. We implemented both NMPC and R²NMPC using the ACADOS library (\cite{Verschueren2021}) and CasADi in Python 3.9. 
For both NMPCs, we employed the SQP RTI as the NLP Solver and HPIPM QP Solver, with a maximum limit of 50 iterations.\\
We derive the reference trajectory for R²NMPC on the Monteblanco racetrack, an actual racing circuit, using a minimum curvature optimization method derived from the global race trajectory optimization framework (\cite{heilmeier2019minimum}). This methodology enables us to compute the optimal trajectory while adhering to specified limits and accounting for the full vehicle dynamics potential, allowing for a comprehensive analysis of the controller's performance near its operational constraints.
Our calculations consider the restrictions imposed by the vehicle interface, outlined in equations (\ref{eq: acc potential}-\ref{eq:hard constraints}). Notably, our TUM research vehicle interface imposes a maximum velocity limit of $37.5 \unit{\meter\per\second}$. In Figure \ref{fig:racetrack}, we provide an illustration of the Monteblanco racetrack layout, the optimal racing line, and the reference velocity profile tailored to the TUM research vehicle.
To model substantial disturbances, we introduce ellipsoid uniform noise samples $\boldsymbol{w} \in \mathcal{E}(\boldsymbol{W})$, which are then added to the measured states:
\begin{equation}
    \begin{aligned}
    \boldsymbol{W}^{\text{sim}}=& \text{diag}(\sigma_x,\sigma_y,\sigma_{\psi},\sigma_{vlon},\sigma_{vlat},\sigma_{\dot{\psi}},\sigma_{\delta_f}) \\ =& \text{diag}(0.8\unit{\meter},0.8\unit{\meter},0.1\unit{\radian},1.1\unit{\meter\per\second}, \\ & 0.2\unit{\meter\per\second},0.05\unit{\radian\per\second},0.01\unit{\radian})
    \end{aligned}
\label{eq:sim_dist_normal_setup}
\end{equation}
For a fair comparison, both R²NMPC and NMPC experience identical disturbance realizations, i.e. $\boldsymbol{w}_t^{\text{R²NMPC}} = \boldsymbol{w}_t^{\text{NMPC}}, \space \forall t \in \{0,\ldots,N\}$.
In addition, we use a straightforward moving average filter to enhance the smoothness of input signals. The filter applied different window sizes to each input state, specifically: $[1,1,4,2,2,3,4,2]^T$.\\
\subsection{Trajectory Following Performance}
\label{subsec:traj_foll_performance}
\begin{figure}[ht]
\includegraphics[width=1\columnwidth]{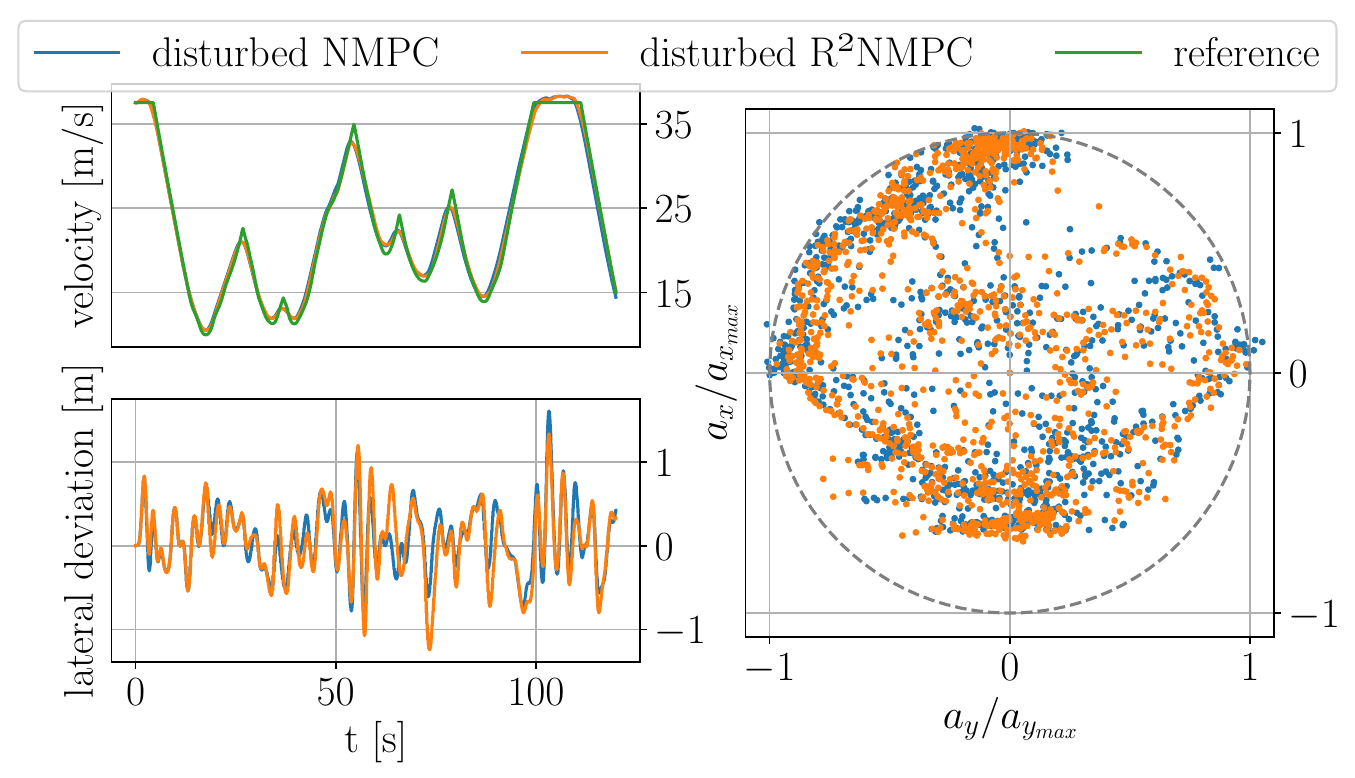}
\caption{Closed-loop performance subject to strong disturbances in Monteblanco racetrack.}
\label{fig:benchmark_disturbed_nmpcs_gg}
\end{figure}

Figure \ref{fig:benchmark_disturbed_nmpcs_gg} presents a comparative analysis of the nominal NMPC and R²NMPC under the influence of disturbances as described in Eq.\ref{eq:sim_dist_normal_setup}.
The gg-diagram in the acceleration plot showcases the contrast in performance between R²NMPC and the nominal NMPC. While R²NMPC operates well within the system limits, the nominal NMPC consistently violates system constraints, as evidenced by the upper center, right, and left portions of the gg-diagram.
Over the course of $N_\text{sim}= 120\unit{\second}  / T_\text{sim}=120\unit{\second} /0.02\unit{\second} = 6000$ steps, the nominal NMPC violates the nonlinear constraints 96 times, i.e. $1.6\%$ of the time. In contrast, our R²NMPC approach demonstrates a remarkable robustness behavior improvement by violating the nonlinear constraints only 17 times, i.e. $0.28\%$ of the time, or \textbf{83\%} improvement over the nominal NMPC. This significant disparity underscores the R²NMPC's ability to effectively account for uncertainties and propagate them.\\
In the velocity plot, both controllers closely track the reference velocity, exhibiting similar behavior. However, a substantial performance contrast becomes apparent in the lateral deviation plot.

\begin{figure}[h]
\includegraphics[width=1\columnwidth]{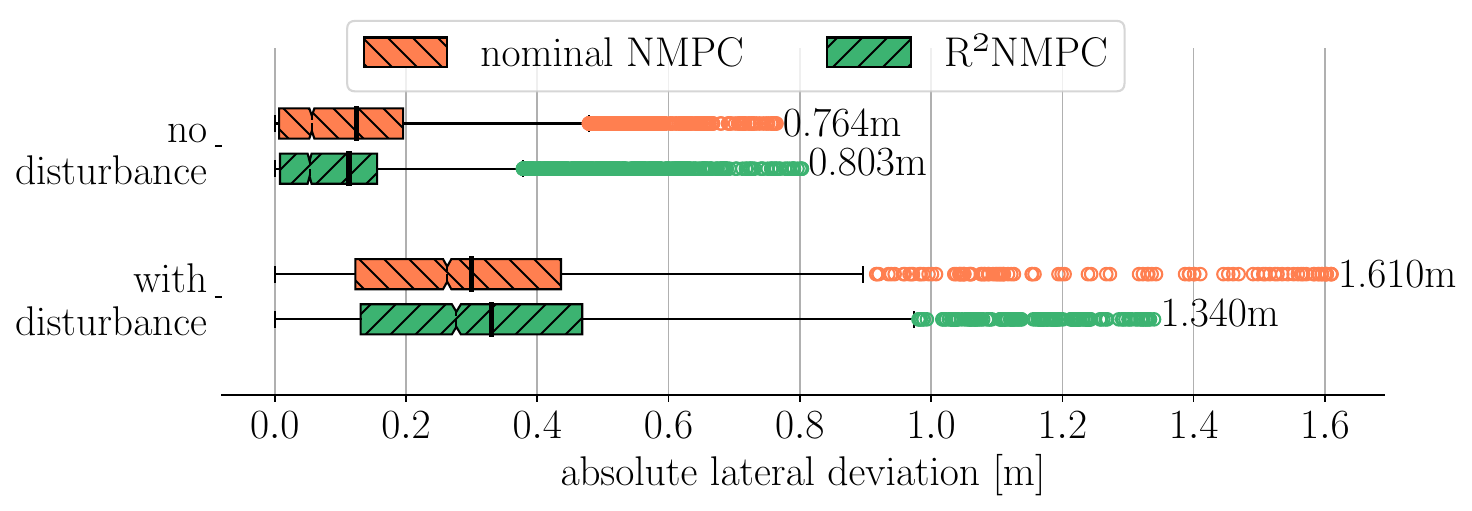}
\caption{Effect of the disturbance on the NMPC and R²NMPC w.r.t. the absolute lateral deviation.}
\label{fig:RNMPCvsNMPC}
\end{figure}

In Figure \ref{fig:RNMPCvsNMPC}, the focus is placed on the lateral behavior of the two controllers, with benchmarking conducted in both disturbance-free and disturbed scenarios.
When assessing the controllers under normal operating conditions, i.e., without disturbances, R²NMPC displays slightly better performance in terms of mean and 75th percentile metrics. However, it's noteworthy that the nominal NMPC outperforms R²NMPC in minimizing the maximum lateral deviation. \\
However, the introduction of disturbances into the system significantly degrades the performance of both NMPC methods. The nominal NMPC maintains a slight edge in terms of the 25th, mean, and 75th percentiles of lateral deviation. Nevertheless, this advantage diminishes when considering the maximum deviation. The absolute maximum lateral deviation signifies the most critical situation in which the vehicle could deviate farthest from its intended trajectory. Minimizing this deviation is important  mitigating the risk of collisions with other vehicles, obstacles, or pedestrians.\\
Comparatively, while the nominal NMPC reaches a peak deviation of $1.61\unit{\meter}$, the R²NMPC significantly outperforms it with a notably lower value of $1.34\unit{\meter}$. This signifies an improvement of approximately \textbf{17\%} in limiting the maximum lateral deviation, highlighting the superior performance of R²NMPC in challenging scenarios.
\subsection{Benchmark of several Robustified NMPCs}
In this section, we conduct a comparative analysis of our R²NMPC approach in relation to the original Robustified NMPC approach outlined in Problem 2 (Eq. \ref{eq:original RNMPC problem}), along with two other Robust NMPC methods. \\
The work by \cite{zanelli_zoro} introduces a zero-order robust optimization (ZoRo) algorithm aimed at approximating the Robustified NMPC in Problem 2. This approach primarily involves simplifying the optimization problem by neglecting the first-order derivatives of the ellipsoidal uncertainty dynamics, resulting in a reduction of the number of optimization variables. Furthermore, Zanelli's work handels the needed approximations and linearizations of the single QP subproblems when solving with sequential quadratic programming. \\
The second work by \cite{kohler_tube} introduces a novel and straightforward constraint tightening technique based on the concept of growing tubes with parameters including a constraint tightening factor denoted as $\epsilon$ and an exponential decay rate represented by $\rho$. This approach allows for easy implementation without requiring significant modifications to the nominal NMPC, thereby avoiding an increase in computational burden or the number of optimization variables.
For the sake of clarity and reference, we refer to the approach proposed by \cite{kohler_tube} as Tube NMPC and that by \cite{zanelli_zoro} as ZoRo NMPC. In our experiments, we compare the performance of the Robustified NMPC (Problem 2), ZoRo NMPC, and Tube NMPC applied in the context of motion control. \\
For the Robustified NMPC and ZoRo NMPC, we use the identical configurations as in Section \ref{subsec:R2NMPC_configuration}. In the case of Tube NMPC, we set $\rho = 62\%$ and $\epsilon = 10^{-3}$.\\
To ensure equitable benchmarking, all experiments are conducted in an identical environment, adhering to the same conditions outlined in Section \ref{subsec:simulation_setup}.
\subsubsection{Closed-loop performance}
\begin{figure}[h]
\includegraphics[width=1\columnwidth]{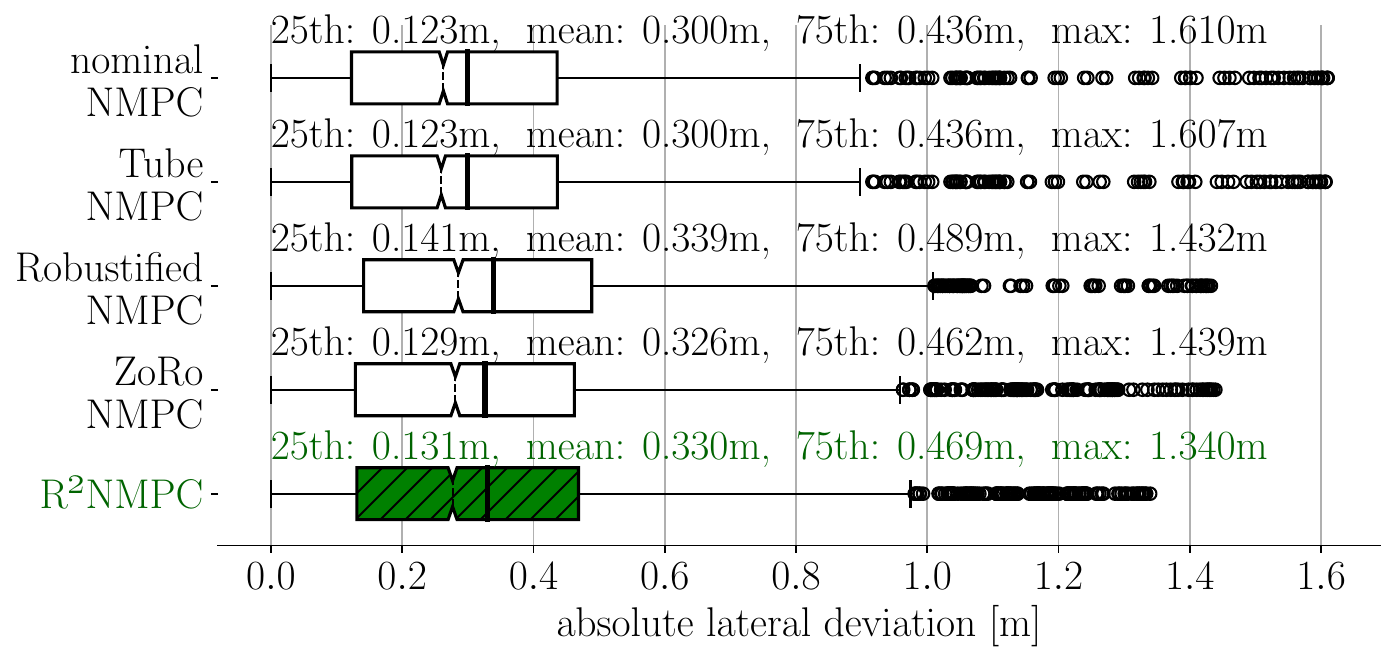}
\caption{Assessing the absolute lateral deviation performance of various Robust NMPC approaches over a 120s duration, equivalent to 6000 simulation steps, in the presence of severe disturbances.}
\label{fig:benchmarkRNMPCs}
\end{figure}

In this benchmark, we evaluate the closed-loop performance of various Robust NMPC approaches in the context of absolute lateral deviation, a critical metric for motion control. All controllers are subject to the same disturbance realization while controlling the vehicle for $120 \unit{\second}$ on the Monteblanco racetrack. The boxplot \ref{fig:benchmarkRNMPCs} delineates that overall, the nominal NMPC and the Tube NMPC deliver a very similar closed-loop performance, but both are surpassed by the 3 Robustified versions. Our R²NMPC maintains similar performance to Robustified NMPC and ZoRo NMPC in the mean and 75th percentile, and performs significantly better in extreme conditions with substantially lower maximum deviations, making it a robust choice for motion control subject to disturbances.
\subsubsection{Constraint satisfaction}
\begin{figure}[h]
\includegraphics[width=1\columnwidth]{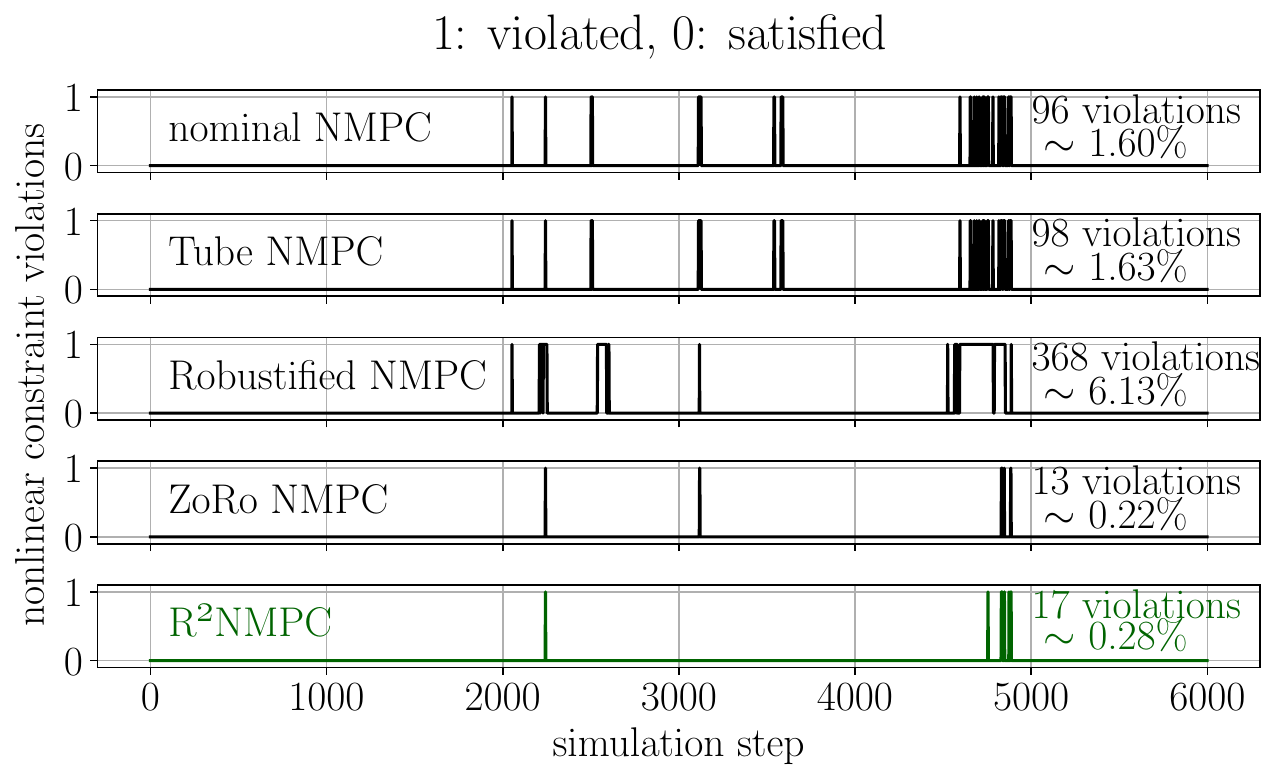}
\caption{Evaluating the nonlinear constraint satisfaction status of various Robust NMPC methods over 120 seconds, equivalent to 6000 simulation steps, in the presence of substantial disturbances.}
\label{fig:constraint_satisfaction}
\end{figure}

In Figure \ref{fig:constraint_satisfaction}, we present the status of the nonlinear acceleration constraints' satisfaction using a binary representation: '1' for constraint violation and '0' for constraint satisfaction. \\
We notice that the nominal and Tube NMPCs exhibit remarkably similar constraint behavior, violating constraints at nearly identical time intervals when subjected to the same disturbance realization. The Robustified NMPC also experiences constraint violations at similar times as the Tube and nominal NMPCs. However, it tends to remain above the constraints for more consecutive steps, distinguishing it as the least effective in handling constraints and achieving a constraint satisfaction rate of only $93.8\%$.
The ZoRo NMPC with only 13 violations delivers the best performance, followed closely by our R²NMPC approach with 17 violations, i.e. demonstrating a nonlinear constraints satisfaction rate of \textbf{99.78\%} and \textbf{99.72\%} respectively.
It's good to note that the controllers were subject to strong disturbances, configured according to Eq.\ref{eq:sim_dist_normal_setup}. 
We observe that our R²NMPC does not violate the constraints on certain occasions, where all other 4 controllers violate the constraints at least once, e.g. between 3000th and 4000th simulation steps.
\subsubsection{Computational times}
\begin{figure}[h]
\includegraphics[width=1\columnwidth]{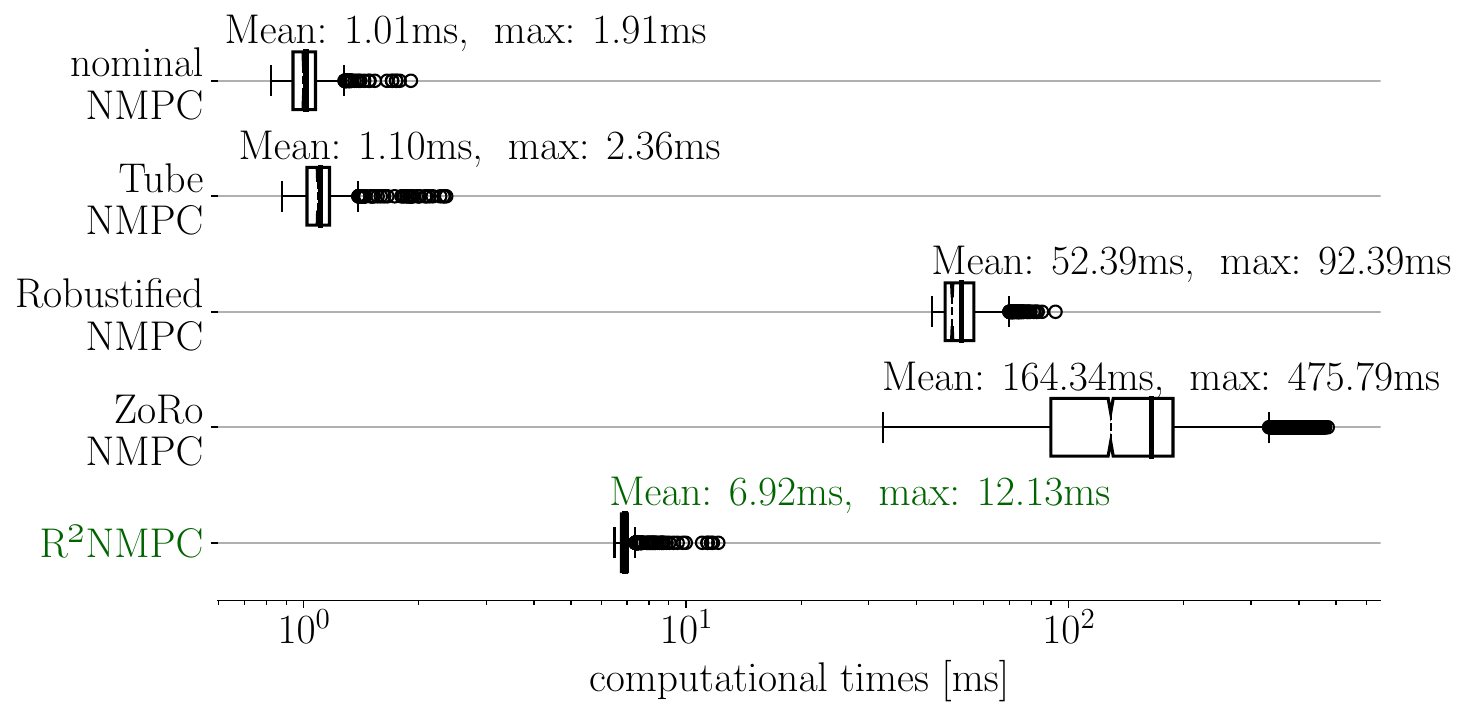}
\caption{Computational times for solving the OCP over 6000 simulation steps.}
\label{fig:computational_times}
\end{figure}

The computational times required to solve the optimal control problem, as illustrated in Figure \ref{fig:computational_times}, offer valuable insights into the efficiency and responsiveness of the five controllers. Notably, the nominal and Tube NMPC controllers exhibit remarkable computational efficiency, capable of solving problems at a minimum frequency of $523.56 \unit{\hertz}$ and $423.72 \unit{\hertz}$, respectively. In contrast, due to the complexity of uncertainty propagation, the Robustified NMPC controller imposes a significantly higher computational load, achieving a minimum problem-solving frequency of only $10.82 \unit{\hertz}$ —more than 48 times slower than the nominal NMPC.\\ 
The ZoRo NMPC controller exhibits the most substantial computational load, with a minimum problem-solving frequency of only $2.1\unit{\hertz}$—almost 250 times slower than the nominal NMPC and at least 5 times slower than the Robustified NMPC. It is worth noting that the work by \cite{zanelli_zoro} suggests that the ZoRo NMPC formulation requires less computational time than the Robustified NMPC. However, this assessment only considers the solver's computational time, overlooking the additional time required to condition the solver at each node in each NMPC step, including ellipsoidal uncertainty propagation and computation of the approximated back-off terms for constraint robustification. We want to emphasize that our evaluation accounts for all these effects in determining the computational time needed for Robustified, ZoRo, and our R²-NMPC implementations.\\
As our R²-NMPC demonstrates performance equivalent to the Robustified and ZoRo NMPC controllers in terms of closed-loop performance and constraint satisfaction, the computational times required for solving the Robustified OCP underscore the significant advantage of our novel R²-NMPC approach, with a minimum problem-solving frequency of only \textbf{82.44Hz}—\textbf{7.6} and \textbf{39} times faster than the original Robustified and ZoRo NMPC, respectively. It's also noteworthy that our R²NMPC maintains a mean solving frequency of $144.5\unit{\hertz}$.
Despite the increased computational complexity associated with R²NMPC in comparison to the nominal approach, our proposed methodology  aligns with our real-time control prerequisites, requiring at least $50 \unit{\hertz}$ update frequency set by the vehicle control interface, thereby underscoring the practical feasibility of our approach in the context of the trajectory following use-case.

\section{Conclusions and Future Work}

We presented a novel Reduced Robustified NMPC \\(R²NMPC) framework, employing ellipsoidal uncertainty sets, that significantly enhances computational efficiency compared to other Robustified NMPC methods, all while maintaining an equivalent closed-loop performance and ensuring constraints satisfaction. Our approach achieves this by dimensionally reducing the optimal control problem (OCP) formulation. We eliminate decision variables related to uncertainty propagation, approximating ellipsoidal uncertainty dynamics based on the last R²NMPC solution and leveraging the benefits of the feedback loop and minimize computational load. Similarly, we approximate the robustification term for nonlinear constraints, i.e., the back-off term, using gradients obtained from the previous R²NMPC solution.\\
The R²NMPC's mean solving frequency of $144.5\unit{\hertz}$ underscores its real-time motion control capabilities for a high-dimensional, highly nonlinear dynamics system subject to strong disturbances, as exemplified by an autonomous passenger vehicle following an optimal race line challenging the vehicle to its full limits. Experimental results demonstrate a 99.72\% satisfaction rate for nonlinear constraints and, notably, a closed-loop performance equivalent to the original Robustified NMPC and ZoRo NMPC \cite{zanelli_zoro} under severe disturbances, all achieved at computational times 7.6 and 39 times faster, respectively.\\
A shared limitation observed in conventional Robust NMPC methods, related to the assumed uncertainty matrix, significantly impacts the closed-loop performance of our approach. To address this, we recommend exploring online estimation techniques for the uncertainty matrix in future work. Additionally, our current assumption of minimal differences between consecutive R²NMPC tasks needs further investigation. Future research should focus on analyzing the performance of R²NMPC in the face of sudden reference changes.


\bibliography{ifacconf}             

\begin{thebibliography}{23}
\providecommand{\natexlab}[1]{#1}
\providecommand{\url}[1]{\texttt{#1}}
\providecommand{\urlprefix}{URL }
\expandafter\ifx\csname urlstyle\endcsname\relax
  \providecommand{\doi}[1]{doi:\discretionary{}{}{}#1}\else
  \providecommand{\doi}{doi:\discretionary{}{}{}\begingroup \urlstyle{rm}\Url}\fi

\bibitem[{Behr et~al.(2019)Behr, Benner, and Heiland}]{behr2019solution}
Behr, M., Benner, P., and Heiland, J. (2019).
\newblock Solution formulas for differential sylvester and lyapunov equations.
\newblock \emph{Calcolo}, 56(4), 51.

\bibitem[{Chen et~al.(2007)Chen, Gao, Wang, and Findeisen}]{chen2007disturbance}
Chen, H., Gao, X., Wang, H., and Findeisen, R. (2007).
\newblock On disturbance attenuation of nonlinear moving horizon control.
\newblock \emph{Assessment and future directions of nonlinear model predictive control}, 283--294.

\bibitem[{Chen et~al.(1997)Chen, Scherer, and Allgower}]{chen1997game}
Chen, H., Scherer, C.W., and Allgower, F. (1997).
\newblock A game theoretic approach to nonlinear robust receding horizon control of constrained systems.
\newblock In \emph{Proceedings of the 1997 American Control Conference (Cat. No. 97CH36041)}, volume~5, 3073--3077. IEEE.

\bibitem[{Han et~al.(2012)Han, Li, and Peng}]{han2012lyapunov}
Han, Z., Li, S., and Peng, S. (2012).
\newblock Lyapunov equations approach for robust nonlinear optimal control problems.
\newblock \emph{Research Journal of Applied Sciences, Engineering and Technology}, 4(13), 2017--2023.

\bibitem[{Heilmeier et~al.(2019)Heilmeier, Wischnewski, Hermansdorfer, Betz, Lienkamp, and Lohmann}]{heilmeier2019minimum}
Heilmeier, A., Wischnewski, A., Hermansdorfer, L., Betz, J., Lienkamp, M., and Lohmann, B. (2019).
\newblock Minimum curvature trajectory planning and control for an autonomous race car.
\newblock \emph{Vehicle System Dynamics}.

\bibitem[{Houska and Diehl(2009)}]{robust-mpc-original-paper}
Houska, B. and Diehl, M. (2009).
\newblock Robust nonlinear optimal control of dynamic systems with affine uncertainties.
\newblock In \emph{In Proceedings of the 48th Conference on Decision and Control, Shanghai, China}, 2274–2279.

\bibitem[{Houska(2011)}]{phdthesis}
Houska, B. (2011).
\newblock \emph{Robust Optimization of Dynamic Systems}.
\newblock Ph.D. thesis.

\bibitem[{Karle(2023)}]{karle2023edgar}
Karle, E. (2023).
\newblock Edgar: An autonomous driving research platform -- from feature development to real-world application.

\bibitem[{Kurzhanski and V{\'a}lyi(1997)}]{kurzhanski1997ellipsoidal}
Kurzhanski, A. and V{\'a}lyi, I. (1997).
\newblock \emph{Ellipsoidal calculus for estimation and control}.
\newblock Springer.

\bibitem[{Köhler et~al.(2018)Köhler, Müller, and Allgöwer}]{kohler_tube}
Köhler, J., Müller, M.A., and Allgöwer, F. (2018).
\newblock A novel constraint tightening approach for nonlinear robust model predictive control.
\newblock In \emph{2018 Annual American Control Conference (ACC)}, 728--734.

\bibitem[{Lazar et~al.(2008)Lazar, De~La~Pena, Heemels, and Alamo}]{lazar2008input}
Lazar, M., De~La~Pena, D.M., Heemels, W.M.H., and Alamo, T. (2008).
\newblock On input-to-state stability of min--max nonlinear model predictive control.
\newblock \emph{Systems \& Control Letters}, 57(1), 39--48.

\bibitem[{Mayne et~al.(2011)Mayne, Kerrigan, Van~Wyk, and Falugi}]{mayne2011tube}
Mayne, D.Q., Kerrigan, E.C., Van~Wyk, E., and Falugi, P. (2011).
\newblock Tube-based robust nonlinear model predictive control.
\newblock \emph{International journal of robust and nonlinear control}, 21(11), 1341--1353.

\bibitem[{Oku and Aihara(2018)}]{lyapunov_equation_2}
Oku, M. and Aihara, K. (2018).
\newblock On the covariance matrix of the stationary distribution of a noisy dynamical system.
\newblock \emph{Nonlinear Theory and Its Applications, IEICE}, 9, 166--184.
\newblock \doi{10.1587/nolta.9.166}.

\bibitem[{Pacejka and Besselink(1997)}]{pacejka1997magic}
Pacejka, H. and Besselink, I. (1997).
\newblock Magic formula tyre model with transient properties.
\newblock \emph{Vehicle system dynamics}, 27, 234--249.

\bibitem[{Raimondo et~al.(2009)Raimondo, Limon, Lazar, Magni, and ndez Camacho}]{raimondo2009min}
Raimondo, D.M., Limon, D., Lazar, M., Magni, L., and ndez Camacho, E.F. (2009).
\newblock Min-max model predictive control of nonlinear systems: A unifying overview on stability.
\newblock \emph{European Journal of Control}, 15(1), 5--21.

\bibitem[{Raji et~al.(2022)Raji, Liniger, Giove, Toschi, Musiu, Morra, Verucchi, Caporale, and Bertogna}]{raji2022motion}
Raji, A., Liniger, A., Giove, A., Toschi, A., Musiu, N., Morra, D., Verucchi, M., Caporale, D., and Bertogna, M. (2022).
\newblock Motion planning and control for multi vehicle autonomous racing at high speeds.
\newblock In \emph{2022 IEEE 25th International Conference on Intelligent Transportation Systems (ITSC)}, 2775--2782. IEEE.

\bibitem[{Rawlings et~al.(2017)Rawlings, Mayne, and Diehl}]{rawlings2017model}
Rawlings, J.B., Mayne, D.Q., and Diehl, M. (2017).
\newblock \emph{Model predictive control: theory, computation, and design}, volume~2.
\newblock Nob Hill Publishing Madison, WI.

\bibitem[{Rubagotti et~al.(2009)Rubagotti, Raimondo, Ferrara, and Magni}]{rubagotti2009robust}
Rubagotti, M., Raimondo, D.M., Ferrara, A., and Magni, L. (2009).
\newblock Robust model predictive control of continuous-time sampled-data nonlinear systems with integral sliding mode.
\newblock In \emph{2009 European Control Conference (ECC)}, 2247--2252. IEEE.

\bibitem[{Smith and Starkey(1995)}]{smith1995effects}
Smith, D.E. and Starkey, J.M. (1995).
\newblock Effects of model complexity on the performance of automated vehicle steering controllers: Model development, validation and comparison.
\newblock \emph{Vehicle system dynamics}, 24(2), 163--181.

\bibitem[{Verschueren et~al.(2021)Verschueren, Frison, Kouzoupis, Frey, van Duijkeren, Zanelli, Novoselnik, Albin, Quirynen, and Diehl}]{Verschueren2021}
Verschueren, R., Frison, G., Kouzoupis, D., Frey, J., van Duijkeren, N., Zanelli, A., Novoselnik, B., Albin, T., Quirynen, R., and Diehl, M. (2021).
\newblock acados -- a modular open-source framework for fast embedded optimal control.
\newblock \emph{Mathematical Programming Computation}.

\bibitem[{Villanueva et~al.(2017)Villanueva, Quirynen, Diehl, Chachuat, and Houska}]{villanueva2017robust}
Villanueva, M.E., Quirynen, R., Diehl, M., Chachuat, B., and Houska, B. (2017).
\newblock Robust mpc via min--max differential inequalities.
\newblock \emph{Automatica}, 77, 311--321.

\bibitem[{Yu et~al.(2013)Yu, Maier, Chen, and Allg{\"o}wer}]{yu2013tube}
Yu, S., Maier, C., Chen, H., and Allg{\"o}wer, F. (2013).
\newblock Tube mpc scheme based on robust control invariant set with application to lipschitz nonlinear systems.
\newblock \emph{Systems \& Control Letters}, 62(2), 194--200.

\bibitem[{Zanelli et~al.(2021)Zanelli, Frey, Messerer, and Diehl}]{zanelli_zoro}
Zanelli, A., Frey, J., Messerer, F., and Diehl, M. (2021).
\newblock Zero-order robust nonlinear model predictive control with ellipsoidal uncertainty sets.
\newblock \emph{IFAC-PapersOnLine}.

\end{thebibliography}
                                                   







\end{document}